# Dhaka Water-logging:
# Causes, Effects and Remedial Policy Options

Hossain Ahmed Taufiq[1]

Conference paper

---

[1] Lecturer (on study leave), Dept. of Global Studies & Governance, Independent University Bangladesh (IUB). Email: taufiq@iub.edu.bd. Paper presented on *'The new megacity: for whom?'* seminar sponsored and organized by Wageningen University, Netherlands. Location: Dept. of Global Studies and Governance, Independent University, Bangladesh– 18 November 2019.

# 1. Introduction:

Dhaka city reeling from extensive water logging during the monsoon (May to October) is now a common scene. Fast, unrestrained and unplanned urbanization have produced water-logging, in turn creating adverse social, economic, physical and environmental impacts on life and living in Dhaka. Traffic disruption, disturbance in normal life, damage of structures and infrastructure, destruction of vegetation and adequate habitats, loss of income potentials are the prime effects of water-logging. Water-logging is also responsible for various forms of waterborne and vector-borne diseases, and the health cost from water-logging is substantial. Recently, a lethal Dengue outbreak battered Dhaka city dwellers (which occurs in urban and suburban locations with higher transmission rates taking place during the rainy season).

This study provides a brief contextual analysis of the Dhaka's geomorphology and natural, as well as storm water drainage systems, before concentrating on the man-made causes and effects of water-logging, ultimately exploring a few remedial measures. The data and information of the contextual brief are mostly secondary. While, the rainfall data was collected from the Bangladesh Meteorological Department (BMD), Dhaka North City Corporation (DNCC), Dhaka South City Corporation (DSCC and Rajdhani Unnayan Kartripakkha: Capital Development Authority (RAJUK) provided the maps and drainage data. In addition to the secondary data, some information was collected through personal interviews and focus group discussions.

# 2. Contextual Brief:

Rivers, waterbodies, and channels tailored the spatial development and livability of the sprawling Dhaka metropolis. Originating from the point of flood free terrace along the river Bouriganga in south, this city's spatial development stretches towards the pre-ostacian Modhupur terrace in the north (Mowla, 2010). The central part of the Dhaka city is developed on the high land with an elevation of 6 to 8 m above mean sea level, and the fringe areas are, located in the flood plains of the Buriganga and Balu rivers with levels of 2 to 6 m above mean sea level (Tawhid, 2004). The fringe areas are constantly flooded. Numerous water channels pass through this terrace, serving as the primary sources of communication. The topographic effect of water and land is visible in the Dhaka's settlement pattern. According to the Atlas of Urban Geology, prepared by the Institute of Water Modelling (IWM, 2003), aside from Buriganga, Balu, and Turag rivers, three more rivers and their branches crisscrossed Dhaka historically:

1. Dulai River (present Dholai channel): presumably a tributary of Balu river, Dulai taking off from an area close to Demra and flowing south-west through Dhaka to join urganga river, the present abandoned channels along Dhaka Medical College-Ramna Park-Segunbagicha are the remnants of the rover.
2. Pandu River: its probable course of flow was through the Mohakhali-Agargaon- Kalyanpur area.
3. Carevan (Karwan) River: it probably aligned along the Begunbari khal-Green Road- Kalabagan- Dhanmondi lake to Turag river.

Now diminished, all three rivers once were inter-connected and contributed towards Dhaka's topography and natural drainage.

## 2.1 Climate and Rainfall:

The Köppen climate classification classifies Dhaka' climate as tropical wet and dry climate. During cold days, Dhaka's temperature drops to 8 °C (46 °F) or less, and during the hot season, it reaches to 40°C (104°F) or more. The annual average rainfall of the city is 1,854 millimeters (Weatherbase, 2019). Three sources are responsible for Dhaka's rainfall: i) the summer monsoon, ii) Nor'westers: the early summer thunderstorms, iii) the western depression of winter (Tawhid, 2004). From April till October, Dhaka's weather is hot and humid, while from November to February, it is cool and dry. Bangladesh Meteorological Department's (2019) five-year rainfall data indicates that around 90 per cent rainfall happens during the hot and humid seasons. During this period, heavy rainfalls, often extends to several days, and total annual rainy days vary from 95 to 131 days (Tawhid, 2004).

*Table 1: Dhaka Rainfall, 2014-18 (Bangladesh Meteorological Department, 2019)*

| Year | | Jan. | Feb. | Mar. | Apr. | May. | Jun. | Jul. | Aug. | Spt. | Oct. | Nov. | Dec. | Annual |
|---|---|---|---|---|---|---|---|---|---|---|---|---|---|---|
| **2014** | Max | 0 | 8 | 10 | 54 | 34 | 52 | 69 | 75 | 37 | 37 | 0 | 0 | 75 |
| | Min | 0 | 1 | 10 | 1 | 1 | 1 | 1 | 1 | 1 | 1 | 0 | 0 | 1 |
| | Total | 0 | 12 | 10 | 80 | 147 | 342 | 212 | 391 | 156 | 49 | 0 | 0 | 1399 |
| **2015** | Max | 2 | 11 | 4 | 58 | 46 | 83 | 90 | 88 | 64 | 18 | 0 | 1 | 90 |
| | Min | 1 | 2 | 4 | 1 | 1 | 1 | 1 | 1 | 2 | 1 | 0 | 1 | 1 |
| | Total | 3 | 17 | 4 | 166 | 185 | 375 | 623 | 395 | 346 | 51 | 0 | 1 | 2166 |
| **2016** | Max | 3 | 8 | 19 | 28 | 52 | 75 | 58 | 33 | 27 | 46 | 21 | 0 | 75 |
| | Min | 3 | 1 | 1 | 2 | 1 | 1 | 1 | 1 | 1 | 1 | 1 | 0 | 1 |
| | Total | 3 | 13 | 55 | 55 | 212 | 212 | 405 | 171 | 138 | 76 | 25 | 0 | 1365 |
| **2017** | Max | 0 | 2 | 38 | 55 | 90 | 139 | 103 | 126 | 94 | 149 | 4 | 15 | 149 |
| | Min | 0 | 2 | 1 | 1 | 1 | 1 | 1 | 1 | 1 | 1 | 2 | 4 | 1 |
| | Total | 0 | 2 | 100 | 228 | 188 | 414 | 584 | 544 | 381 | 412 | 6 | 33 | 2892 |
| **2018** | Max | 0 | 16 | 2 | 77 | 52 | 82 | 60 | 25 | 23 | 24 | 13 | 12 | 82 |
| | Min | 0 | 4 | 1 | 2 | 1 | 1 | 1 | 1 | 1 | 1 | 13 | 1 | 1 |
| | Total | 0 | 20 | 3 | 309 | 392 | 366 | 354 | 141 | 76 | 45 | 13 | 13 | 1732 |

## 2.2 Natural Drainage System:

Dhaka's natural drainage system includes water retention bodies and channels (commonly referred to as 'Khals'). Connected to Dhaka's adjacent rivers, these khals discharge the monsoon rainfall-runoff, accumulated in the retention bodies. There are 40 khals in Dhaka, exceeding 90 kilometers in length, and their individual catchment area varies 6 to 40 sq. km (Chowdhury et al., 1998). These khals drain approximately 80% of Dhaka's water (Mowla, 2013).

Table 2: Dhaka's Canals (Tawhid, 2004)

| Name of the khals | Catchments Area (Sq. km) | Length |
|---|---|---|
| Begunbari khal | 37.7 | 6.5 |
| Dholai khal | 16.8 | 4.0 |
| Segunbagicha khal | 8.3 | 3.5 |
| Gerani khal | 6.7 | 3.4 |

All these khals have/had their outlets to the Buriganga, the Sitalakkhya, the Balu and the Turag rivers, which were inter-connected (Mowla, 2013).

Apart from the khals, there are also many water storage areas such as lakes and low laying lands which function as retention areas. According to Bangladesh University of Engineering and Technology (BUET) (2009), Dhaka's water bodies can be categorized into six types – ditch, natural depression / lowland, pond, natural depression / lowland, lake, swamp/wetland.

Table 3: Categorization of in land water bodies in Dhaka (BUET, 2009)

| Ditch | Pond | Lake | Natural Depression | Khal (Canal/Channel) | Swamp |
|---|---|---|---|---|---|
| 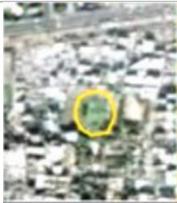 | 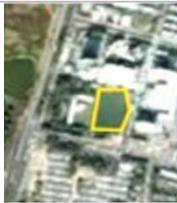 | 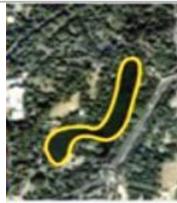 | 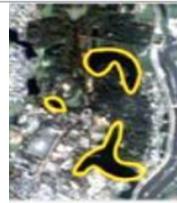 | 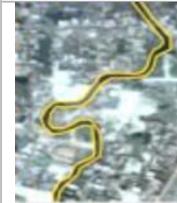 | 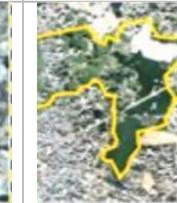 |
| Small, temporary, and perennial accumulation of water. People dump garbage in it. | Perennial waterbody commonly dug with every rural homestead. Big ponds are called "dighi" | Man made waterbody, commonly dug in urban areas. | Single waterbody in wet season, split up into multiple pockets in dry season | Man made connection with river, beel (a lake-like wetland with static water), floodplain etc. | Track of low-lying land generally formed up due to the filling of the lake basin. |

Currently, there are four major lakes in Dhaka, the details are given in the table:

*Table 4: Dhaka's Lakes (Tawhid, 2004)*

| Lakes | Length (meter) | Area (square kilometer) |
|---|---|---|
| Gulshan lake | 3800 | 0.480 |
| Dhanmondi lake | 2400 | 0.167 |
| Crescent lake | 650 | 0.016 |
| Ramna lake | 400 | 0.020 |

Apart from the above four lakes, Dhaka also has a lakefront in Hatirjheel-Begunbari areas, which is commonly referred to as the Hatirjheel lake. The government has spent 19.71 billion takas for the renovation of this 1.22 square kilometer water body (Daily Asian Age, September).

## 2.3 Rainfall Run-Off, Backwater Effect and Retention Storage:

The rainfall run-off is discharged to Dhaka's adjoining rivers. Most Buriganga distributaries during the monsoon remains high, causing backwater effect in the connected drainage system (the storm sewers and channels). For several days, the flow velocity in the system remain sluggish as the flood-water passes through the rivers. Luckily, Dhaka's lakes and low-lying areas work as large retention, saving the city flooding during heavy storms. But, increasing human activities are threatening these retention bodies, causing water-logging.

## 2.4 Stormwater Drainage System:

Dhaka's first drainage master plan was prepared in the 1960s and revised in 1991 with the support of JICA. Later in 2011 a new study was carried out with the support of World Bank. Dhaka Water Supply and Sewerage Authority (DWASA) is responsible for water supply and drainage of Dhaka city. Drainage is managed through two separate sewer systems: one for drainage of domestic wastewater and the other for drainage of storm water. The research topic is confined to the storm water drainage system. Based on the physiological nature of Dhaka, DWASA (2012) divided the city drainage basin has been divided into the following three regions:

1. **Dhaka-Narayanganj-Demra (DND) area:** Bangladesh Water Development Board (BWDB) first developed the area as part of an irrigation project. Protected from flood by polders, this 60 sq. km comparatively little economic value, urban fringe land has undergone tremendous pressure for development despite general embargo by RAJUK's embargo.

2. **Western Dhaka:** This area is flood protected, and the total protected area is around 142 sq. km extending from Tongi Khal (north) to DND area (south) and Turag-Buriganga River (west) to Railway line-Pragati Sarani-Atish Dipankar Road (east). These catchments include the sewerage system which typically receive the discharges of household and industrial waste water. DWASA (2012), however, acknowledges that the existing sewerage system is not properly functional.

3. **Eastern Dhaka:** Much of Eastern Dhaka's 116 sq. km area is unprotected from flood. Extending from Atish Dipankar Road-Pragati Sarani-Railway line in the west to the Balu river in the east and DND area in the south to Tongi Khal in the north, this part of Dhaka lacks a proper sewage disposal system. The drainage network receives the burden of discharging domestic and industrial effluent as well as the stormwater, and typically discharged to the Banani, Gulshan, and Hartijheel lakes. Though previously reserved as a flood plain, residential development overtook the area.

Three storm-water pump stations operate in Dhaka. Located in Goranchanbari, Kallyanpur, and Dholai khal areas, all these stations fall under the western storm-water catchments. Government planned to install two stations in the eastern catchments, and a new pump on Kamalapur had been inaugurated in 2016 (DWASA, 2016). Kallyanpur and Goranchanbari pumps are connected to stormwater retention ponds collecting stormwater (and wastewater discharged to the drainage system) from their respective drainage zones.

## 3. Problem Statement:

In last 40 years or so, unchecked and unplanned urbanization has taken place in Dhaka. The city observed insensitive development of areas through private land developers and real estate business. Thus, Dhaka is experiencing a staggering increase in impervious areas, creating obstruction to natural drainage pattern, and reducing detention basins. These are creating shortened runoff concentration-time and increased the peak flow. As a result, flooding due to rainfall is a severe problem for Dhaka city that remains inundated after each severe shower mainly due to the drainage congestion. The city has experienced waterlogging for the last couple of years which creates large infrastructure problems for Dhaka city that remains inundated after each severe shower mainly due to drainage congestion. Waterlogging creates large infrastructure problems for the city and a huge economic loss in production for the city together with substantial damages of existing property and goods. Besides, ecological balance of the city is also disrupted, and diseases spread which is a gross inconvenience to its inhabitants.

The city is protected from river flooding by an encircled embankment called Buckland Bund and the Western Flood Protection Embankment, further obstructing the natural drainage. During the rainy season (May to October), the high flooding in Dhaka's surrounding rivers create back water effect in the internal drainage system. Enough retention and detention capacity of rainfall-runoff and sustainable drainage development during urban design and planning are now crying needs for Dhaka.

The authorities in Dhaka introduced several crash programs to combat water-logging. In 1968, the Department of Public Health Engineering (DPHE) adopted the first master plan to protect the developed area of 75 km$^2$ from floodwaters and to drain internal storm waters (Mowla, 2005). In 1975, BWDB developed a detailed scheme for an area of 145 km$^2$ as a follow-up of the '68 plan (BWDB, 1990). DPHE in 1976 introduced another scheme for internal drainage. Nevertheless, financial constraints prevented the authorities from approving the proposals, instead DPHE adopted a crash program in 1976 (Tawhid, 2004).

Current storm water drainage is inadequate and cannot cope with recurring floods and storms. In a 1981 study, drainage system improvement was recommended and as a follow-up the "Dhaka Metropolitan integrated Urban development Project" incorporated a framework on the proposed improvement (Tawhid, 2004). Afterward, the Department of Public Health Engineering (DPHE) and later the Dhaka Water and Sewerage Authority (DWASA) constructed about 140km of small to large diameter sub-surface drains and re-excavated approximately 100km of the canal system (DWASA, 2012). Dhaka City Corporation (DCC) also constructed another 120kms of surface and sub-surface drains (DWASA, 2012).

Isolated or whimsical projects cannot solve Dhaka's waterlogging problem. A large amount of money has been spent on such projects in the past without any tangible success. Well-coordinated and comprehensive plan incorporating all the concerned authorities can reduce the waterlogging.

Research questions:

Taking the abovementioned issues into consideration, this paper seeks to answer three research questions:

1. What are the man-made factors and causes of water-logging?
2. What is the state of the existing anti-water-logging infrastructure?
3. What are the social, economic, physical and environmental impact of water-logging in daily life and living?

## 4. Methodology:

Both primary and secondary data were needed to investigate the research questions of the study. Below are the techniques used to collect data and information.

**Focused Group Discussions and Informal Interviews:** To understand the key causes and effects of waterlogging, a qualitative data collection as informal interviews of experts, pubic and elected officials as well as focused group discussions (FGD) with city dwellers has been conducted (lists in the annexure). FGD participants were selected from three extremely waterlogging prone areas through a purposive sampling technique. While the FGD participants shared their waterlogging experiences, the experts and public officials provided the technical information behind the water-logging.

**Maps and Secondary Data:** DNCC Zone 4 office provided a zone map and a report consisting information and maps of *khal*s, water retention ponds and pumping facilities. Bangladesh Meteorological Department (BMD) provided the latest five-year monthly rainfall data. Some maps and information were collected from RAJUK draft structural plan report (2016-2035) and Dhaka Dhaka Water Supply & Sewerage Authority (WASA) Sewerage (2012) and Drainage (2016) master plans. The dengue data was collected from the Directorate General of Health Services (DGHS). For other secondary sources, various scholarly papers, reports and news articles have been reviewed.

# 5. Causes of Water-logging:

According to urban planner, Dr. Md. Shakil Akter (in a personal interview on 16 November 2019), some key causes of water-logging are: "1. Unplanned urbanization, 2. Lack of synchronization of the service organization, 3. dysfunctionality of existing services, 4. untimely waste disposal, 5. inadequate drainage system, 6. illicit encroachment of *khal*s, 7. lack of waterbodies, 8. unsystematic construction, and 9. obstruction of *khal* networks." By taking his points, the study provides an analysis of factors behind water-logging.

**3.1 Urbanization, Encroachment and Disappearance of Natural Drainage System:**

Dhaka's natural drainage systems are disappearing. This disappearance is one of the primary causes behind the water-logging. Overpopulation and rapid development of new residential areas, illicit land filling, encroachments on lakes, and clogged drains due to ill-managed garbage and solid waste are some of the reasons behind the disappearance.

During 1960's, around 50 *khal*s crisscrossed Dhaka City, with the total length of 256 km, but gradual encroachment reduced the number to 26 *khal*s, with only 125km total length (Tawhid, 2004). Dholai *khal*, which used to be one of the principle river-route in oldest part of Dhaka, now almost disappeared due to four decades of wrong policies (Huq and Alam, 2003). Excavated by the Mughal Subedar Islam Khan in 1610, this *khal* had interconnected outlets to the Buriganga, the Balu, the Sitalakkhya, and the Turag rivers (Banglapedia, 2015). Unplanned and irregular road constructions over the canal killed its flow, rendering it to extinction. Dhaka's other *khal*s faced similar fate. According to Mowla and Islam (2013):

> "Segunbagicha Khal extending from Shahbagh to the Jirani Khal via the Manda Bridge… Begunbari Khal extending from Dhanmondi Lake to Trimuhani via Rampura before emptying into the Balu River, the Ibrahimpur canal, the Khathalbagan-Rajarbagh canal, the Gopibagh canal together with other minor canals of the city are all victims" (p.24).

Dhaka requires at-least 20 retention ponds, sizing Hatirjheel to tackle the storm water run-offs (Bangladesh news, 2009 July). On the contrary, around 1,000 ponds have now been destroyed in last few decades (Bangladesh news, 2009 March). Due to illegal encroachment, five rivers surrounding Dhaka including Sitalakkhya and Buriganga lost 324 hectares (Mowla and Islam, 2013). This is a gross violation of the Wetland Protection Act, 2000. Chowdhooree (2010) claims, with the current rate of loss, all of Dhaka's temporary wetlands will disappear by 2031. Situation worsened to such a level that even if the Water Body Conservation Act, 2000 is implemented, the wetland loss cannot be stopped (Sultana, 2007). According to Sultana (2007), due to rapid population growth, Dhaka's fringe areas are developing in 'S' shape, indicating the early stage of urbanization. The danger of such curve development is the disappearance of more wetlands, as the gradual population and economic activities growth means continuous filling up of more wetlands.

Owing to the real state and residential purpose land filling, vast water catchments in Aftabnagar, Badda, Baunia, Meradia, Banashree, Ashulia, Meradia and Amin Bazar areas are dangerously shrinking. These are areas marked as flood flow zones in the Dhaka Metropolitan Development Plan (DMDP), and land development in this zone clearly violates the Wetland Conservation Act, 2000. The rise of impervious surface in these wastelands increased the water-logging hazard that now regularly swamp Dhaka city during the monsoon.

The Dhaka Tribune reporter Anik (2019, July) informs that approximately 3,483 acres (1,410 hectares) of water bodies, low-lying lands filled up over the last nine years.

*Table 5: Dhaka's Total and Filled Up Waterbodies (Anik, 2019)*

| Areas of water bodies and low-lying land filled up in Dhaka City | | | |
|---|---|---|---|
| Type | Total Land (Hectares) | Filled Land (Hectares) | Percentage of Land Filled in |
| Flood Flow Zone | 760 | 436 | 57% |
| Water Retention Area | 1,826 | 628 | 34% |
| Waterbodies | 1,280 | 344 | 27% |
| Total | 3,867 | 1,410 | 36% |
| Areas of water bodies and low-lying land filled up in RAJUK areas | | | |
| Flood Flow Zone | 30,230 | 7,730 | 25% |
| Water Retention Area | 2,238 | 789 | 35% |
| Waterbodies | 8,380 | 603 | 7% |
| Total | 40,848 | 912 | 22% |

## 3.2 Low Road and Area Elevation:

Dhaka WASA identified at least 48 water-logging prone areas (Shafiq, 2018). Though poor drainage and sewerage contributes to the waterlogging in many areas, some of these areas fall in low elevation pockets surrounded by comparatively higher grounds. Bashundhara, Mirpur's Kazipara, Senpara, few pockets in Basabo, Khilgaon, Meradia, Shantinagar, Badda, Postagola, Demra, Amlapara, Siddheswari and parts of the old town fall in such category. Unable to pass through the sewage and drains, rainfall run-off flows down to these areas and create water-logging. During the personal interview, Dhaka Ward-14 Counsellor Humayun Rashid Jonny informed that, "Water drains from high elevation areas to low elevation areas. Senpara, Kazipara and Shewrapa (Mirpur) get waterlogged by water from Mirpur 10, 11 and 12. There is no place for rain-water to drain. For example- before there were *khal*s and many ditches. The parliament has lake, so water drains there but our area does not have any water retention space."

Another problem is the low elevations of the internal/community roads than the connecting main roads. Rain-water flows to this area and creates water-logging. The bad drainage systems do not help the situation and sometimes stuck rainfall water remains for 3-4 days. During Focused Group Discussions in Kazipara (Mirpur), Moulobhirtek (Khilgaon), and Basabo (Sabujbag) *thana* (sub-district) areas, the participants made similar complains.

## 3.3 Poor drainage and infrastructure system:

Due to obstruction, degradation, and demolition of natural drainage systems, Dhaka is prone to frequent flooding. A Detailed Area Plan (DAP, 2007) was developed to address the issue. However, rather than focusing on adequate space for water retention ponds and permeable surfaces like- parks and other unpaved surfaces, DAP focus is on roads and structures, pump stations and embankments approach. This approach risks intensifying residents suffering rather than diminishing them. The geo-morphology of Dhaka city indicates that a reservoir-based gravity drainage system will work better than pump and embankment-based flood control strategy. For instance, Dhanmodi area is encircled by water basin. As a result, this area is less susceptible to flood and waterlogging. On the other hand, Amlapara or BUET campus or Siddheswari area more susceptible to flood and waterlogging for the same reason that there is no place for the water to flow to.

According to an official of DNCC, Dhaka norths' waste and storm water flows to the nearby Turag river. Following flow diagram and details obtained from DNCC show how the system works:

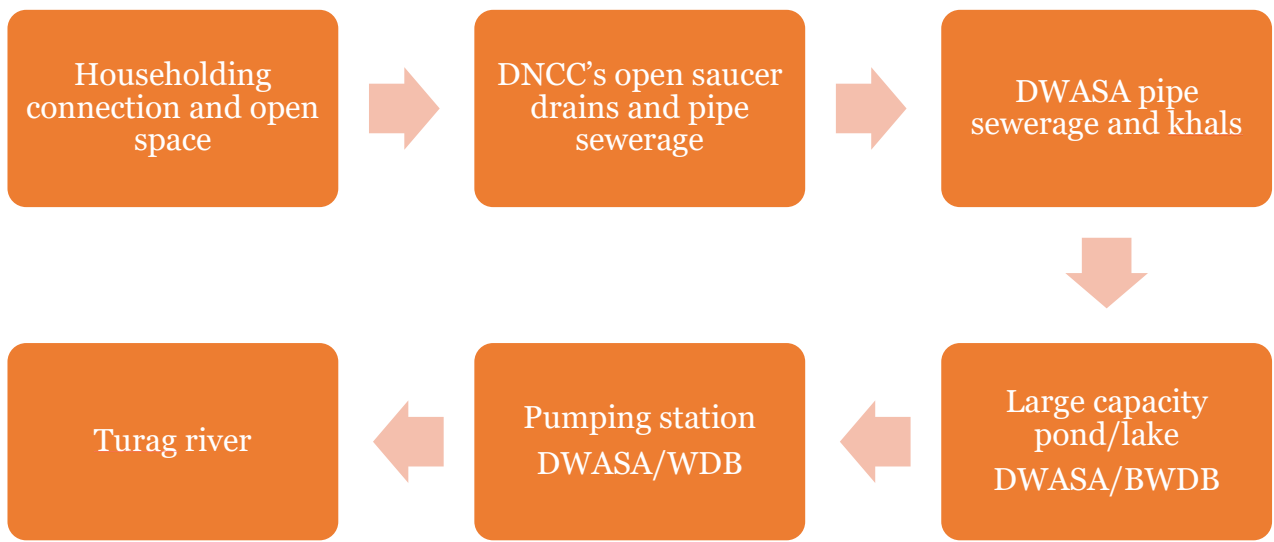

The picture above gives an idea of water flow and the detail is given below-

- **Household connection and open space:** still water created by either rain or everyday usage is known as householding connection and open space water. Household water or outdoor water first go towards the drains.
- **DNCC's open saucer drains and pipe sewerage:** it is usually built while building the roads on both sides of footpaths. Household or rain water come through the pipes and fall into these drains. Though these drains the water pass to the DWASA-controlled *khal*s.
- **DWASA pipe sewerage and *khal*s:** it indicates the *khal*s that are situated in Dhaka city. Water run through pipes or directly into these *khal*s. Through these *khal*s water to go different water bodies or holdings where water is removed through pump.
- **Pumping station:** in the pumping station water is dumped into the river.

In this process water passes from household or open space to drains to the river as the last step and this is the current and only water channel system. However, the system functions only theoretically. FGD in Kazipara, Mirpur (a DNCC area) reveals that, narrow drain-pipes, clogged and filled drainage system by solid waste, garbage, sands, and road side constructions materials, lack of alternative drainage, and poor sewage connections to the *khal*s and big rivers are some of the key reasons behind water-logging. FGD participants also ranted out against concerned authorities poor drain construction:

> "If rain continues for an hour and half or two, water-logging happens in the area. Drainage construction work that takes place here, are extremely poor in quality. For instance, manhole covers which are basically cement slabs broke within two days after their construction."

When asked about the problems, the DNCC Executive Engineer (Zone-4), Mollah Md. Nuruzzaman replied,

> "DNCC does not specifically work on waterlogging. However, it helps eradicate the waterlogging effect. For example, DNCC built sewerage for a better drainage system but it also works to reduce waterlogging."

He blamed DWASA's inadequate sewerage system, narrow drainage line, encroachments in khals and retention ponds, dumping of bricks, sands, cements and other materials in the water bodies, reduction of depth of Turag river for water-logging. Among other reasons mentioned by him are summarized in the below points:

***Low powered pumps:*** Existing pumps are of very low power, thereby they pump out inadequate amount of rainwater.

***Solid waste, garbage and encroachment:*** Owing to the residential development activities, retention ponds and *khal*s connected to the pumps have become subject to land and solid waste filling.

***Pumps do not operate when needed:*** When rain starts, water from all direction starts flowing towards the pumping station area, but pumping does not start at the same time. It starts after the water is full. It does not pump out as per the rate water is flowing in the adjacent area. The result is waterlogging. Poor management or negligence of the employees are the reasons behind this.

***Ignorance:*** According to the DNCC engineer, "people are extremely ignorant." They throw all types of wastes into the drains. They slaughter animals (such as cows) beside the drains as a result bloods flow to the drain; they do not even wash it with water. They throw other parts of the slaughtered into the drains. Construction works also contribute towards blocking of drains. Waste materials from the constructions like broken bricks, sand, cements, etc. are dumped in the drains.

***Lack of maintenance:*** Maintenance is not done where needed. It has been seen that where a drain has become unusable, instead of repairing it, a new one is built over or beside the old one. It does not solve the problem but increases the cost. If old drains are repaired, it would cost less because it is less costly to maintain than building a new one.

***Lack of synchronization:*** The organizations that work in the city, do it separately without any integration. This is another primary reason of waterlogging in Dhaka city. DWASA works on *khal*s in their own way, BWDB on waterbodies or pumping stations or retention ponds in their own way, and city corporation works on drains in their own way. No one communicates with others even though all these works are interconnected. If there is a problem in any place, the entire water flow channel gets affected. For example- if DWASA builds a new *khal* or works on an old one and does not connect it to the drains, the water from the drains will not flow to the *khal*s properly. On the other hand, if BWDB's work is not synchronized with DWASA then water will not flow from *khal*s to the pumping stations.

***Pucca (solid/pitch/permanent) road:*** previously roads were not pucca and there were plenty of open spaces. Rain-water could easily go underground. But now everything is Pucca so water cannot go underground.

In another interview, BUET Professor Shah Jahan Mandal gave similar opinion regarding the pump's functionality and capacity, lack synchronization among authorities and lack of awareness among Dhaka residents. He added (summarized below):

- Drainage system do not properly work. Although the pumping stations located at Kallyanpur, Dholaikhal, and Goranchanbari are functional, but the drainage system linked to these facilities leading to the river are not adequate and often not operational.
- Box Culverts do not properly function, because of deep design flaws. Most of these culverts are shaped in a rectangular format, but they should have been designed in UCEP format.
- Sewerage systems are not properly cleaned and maintained. In fact, Bangladesh do not have the necessary tools and machineries to properly maintain the sewerage system. Unplanned road construction and real estate expansion, inept urban planning, lack of garbage and solid waste management resulting in blockage of Dhaka drainage system.

A recent DSCC and Dhaka WASA joint report found several waterlogging chokepoints in the Dhaka South (Shafiq, 2018). According to the report rainwater from Dhanmondi-8/A Staff Quarters intersection, Dhanmondi-27, Kathal Bagan, Rapa Plaza, Gastro Liver lane, Green Road, Kalabagan Dolphin Lane, Meradia, and Madartek, traverse through the Panthapath Box Culvert, Rampura and Hatirjheel canal, and finally come to pass in the Balu River. Rainwater drain into the Buriganga River through the Buriganga sluice gate from Kazi Alauddin Road, Nazimuddin Road, Hossaini Dalan, New Market, BGB gates 3 and 4, Chawkbazar, west-south Bottola, Lalbagh, and Bangsal. Rainwater from Mouchak, Malibagh, Shantinagar, Motijheel, Dilkusha, Bailey Road, Siddheswari, Circuit House Road, Doinik Bangla, Segunbagicha, Paltan, Shantibagh, Fakirapul Rajarbagh, and Arambagh end up to the Buriganga and Balu Rivers through the Kamalapur Pump, Segunbagicha Box Culvert, Maniknagar canal, Manda canal, and Jirani canal. In Gulistan, there is no drainage, triggering waterlogging Bangabhaban and Secretariat areas after rainfall. Water from Mugda, Basabo and Khilgaon passes through the Madartek, and Basabo, and Trimohoni khals, eventually reaching the Balu and Buriganga rivers. Rainwater from Laxmibazar, Kaptan Bazar, and Agamasi Lane flows through the English Road, Dholaikhal box culvert, ending up at the Sutrapur pump. Rainwater from Postogola, Jurain, Muradpur, Kodomtola, Shyampur, and Doyaganj rail-bridge drains through the Zia Sarani canal, Rosulbagh, and Shimrail pump (Water Development Board), draining into the Shitalakshya river. In Mir Hajaribagh, there is no drainage.

Similar to the Kazipara (a DNCC area) FGD findings, the FGDs in Khilgaon and Bashabo-Sabujbagh areas (DSCC areas) reveal that the narrow drain-pipes, clogged and filled drainage system owing to solid waste and garbage dumping, poor sewage connections to the *khal*s and big rivers as the key reasons behind water-logging. In addition, FGD participants pointed their figure at corruption, lack of coordination and ignorance among the concerned authorities, dumping and littering by ignorant people, ignorance in drain cleaning as the primary causes behind water-logging.

### 3.4 Waste Management System:

Dhaka's haphazard waste management is one of the core reasons behind this city's water-logging. Unprecedented urban congestion, incredible population density, explosive population growth, and unchecked urban migration have made it impossible to clean the drains and streets as fast as the waste thrown onto them. With a density of 47,400 people per square kilometer, Dhaka tops the world's most densely populated city list (Dermographia World Urban Areas, 2018). According to Kazi (2002), this City daily generates daily *3500-4000 tons solid waste*, making the per capita generation about 0.5 kg. In 2016-17, the total waste generated was approximately 8,52,390 tonnes (Khan, 2019). During the rainy season, the amount of waste generation increases, as this is the time when most of the seasonal vegetables and fruits become available into the market (Tawhid, 2019). The left-over of fruits, trashed food, plants and grass, brick, paper, dirt and polythene materials make up majority of Dhaka's waste.

Management of this massive waste primarily lies on the two city corporations—the Dhaka North City Corporation (DNCC) and the Dhaka South City Corporation (DSCC). DSCC's waste dumping ground Matuail Landfill, with an area 40-hectare -- already has 20-meter-high piles of garbage, while DNCC's 21-hectare Aminbazar landfill -- is already nine-meter-high (Khan, 2019). Both city corporations are unable to process the waste using the traditional landfill method. With recent donor support, they have taken several projects such as community-based waste management activities, urban public and environmental health development projects, medical waste recycling plant development of sanitary landfill, and waste-based power plant (Khan, 2019; Ahmed, 2019; Chandan, 2019). But Mahmud (2018) reports that the managements have seen no major improvement due to following reasons:

- **flouting rules and regulations:** Negligence complaints, in duty against DNCC and DSCC staffs and sweepers are common. Although, nearly 4,000 tons of daily household garbage generates in Dhaka, the DNCC and DSCC only report 500 tons or less. Such disparity clearly indicates that citizens complain is not irrelevant. Citizens also do not follow rules, many dump garbage ignoring designated spots and waste bins.

- **unsuccessful mini-bin project:** In 2016, Dhaka's city corporations installed 6,000 waste bins. Lack of strict waste disposal law and Carelessness among Citizen's led the project to go with fiasco. People throw garbage here and there, instead of the bin. In fact, many bins got stolen.

- **incomplete STS projects:** Lack of coordination and blame games are common among the concerned authorities. For instance, DSCC and DNCC officials blame RAJUK's faulty plan for not being able to build secondary transfer stations (STS) in every ward for efficient waste management. In 2013, World Bank funded the STS project. The project missed the December 2015 deadline. According to the officials, lack of free space, illegal occupation in the allotted spot, and interreference by the influential people for the delay and incompletion. As of February 2018, in DNCC, among 72 STS only 51 have been finished, while in DSCC, only 12 among 45 have been finished. Such procrastination and blaming do not help, as unmanaged wastes quickly block the drainage and sewerage systems.

- **no steps to tackle the construction dumps:** Both city corporations lack strict regulations against irregular construction material dumping in footpaths and roads. Unchecked and unprotected open construction and demolition of old buildings, random road excavation and repair often lead to congestion in the drainage system.

- **open garbage trucks:** Mobile garbage removal directives are often not properly followed. Garbage truck drivers and cleaners are supposed to cover the waste when moving them to the landfill from the STS. But most often the truck tops remain open and wastes randomly fall in the roads. Rainfall washes away these dropped wastes to the nearby drains and sewerage systems.

- **fund crisis for waste-based power plants:** DNCC and DSCC's flagship project the 'waste-based power station' never took off. The original construction deadline was 2013, but as of today, Italy-based Management Environment Finance SRL Ltd failed to disburse the funds needed initiate the project.

- **failed 3R waste management:** In 2012, Bangladesh Climate Trust Change Fund provided USD 2.47 million to The Department of Environment, which prescribed the 3R method- reduce, reuse, recycle. Inadequate sensitization among public and repeated poor waste management rendered the project failed.

- **inadequate waste treatment plant:** Matual and Aminbazar landfills produce substantial leachate and have been polluting the nearby environment. Among the two landfills, only Matuail has a leachate treatment plant, but it is insufficient against the heavy load of waste. Leachate, a liquid that drains or leaches from a landfill, is hazardous to arable lands, water resources and aquatic lives.

Among the solid waste, most dangerous is plastic and polythene. The government took initiatives to ban regular use of polythene in 2002. The ban's effect did not last long. After a short recess polythene again become available in the market. Cheap and readily available, most people like to reuse polyethene bags. One of the many hazards of polythene and plastic is that careless littering of these items blocks drainage systems. A recent report published by Earth Day Network (2018) ranked Bangladesh as world's 10th most plastic polluting countries. The estimated contribution of plastic is eight percent of the country's total annual waste, of which around 200,000 tonnes go into the ocean and rivers (Seppo, 2018). According to former city mayor Mohammad Hanif, "Indiscriminate dumping of polythene bags has been creating serious environmental hazards and water logging because this insoluble object is choking the drainage system and causing overflow of filthy sewerage water" (Islam, 1998).

## 6. Effects of Water Logging

Waterlogging creates varied physical, social, economic and environmental problems in the urban life:

**4.1 Physical Problems:**

Because of the water-logging, residential houses, schools, colleges, and other buildings in the low laying areas go under water. Most of these buildings are made of bricks. As the water-logging creates corrosive after effect such as dampness and salinity, the sturdiness of the brick foundation deteriorates. Most vulnerable are the low-income people who live in slums, as these settlements receive the hardest hit. Most often these people cannot rebuild their house and forced to migrate in other areas, resulting in intense economic and psychological hardship.

Dampness, heave, and subsidence during and immediately after the water-logging damage road pitches and create dangerous pothole in the road. Underground utilities such as metalloid pipes used for water, sewerage, telephone, etc. purpose receive damage and lose durability.

**4.2 Social Problems:**

One of the irksome effects of Dhaka's water-logging is the disruption to traffic. Over 25mm rainfall in Dhaka can create 20 cm or higher trapped rainfall run-off in the roads for hours, forming puddles (Tawhid 2004). The storm water drainage cannot drain out this water, making Dhaka's awful traffic jam even more pathetic.

Despite few botherations, affluent people can employ various coping strategies against water-logging. On the contrary, poor people's suffering knows no bound during water-logging. As majority of them live in unstable sites, they are forced to stomach the brunt of poor drainage. Bad drainage allow flood to enter their house and the surroundings. Often, the water stays for few days or weeks, polluting supply water and aquatic assets, breeding mosquito and other vectors, leading to health problems, direct and indirect financial loss, and other livelihood damages.

**4.3 Environmental and Health Problems:**

Though, theoretically Dhaka WASA has two separate sewer systems, but in reality, bulk of the domestic wastewater passes through the storm sewer, falling into and polluting the receiving water bodies.

One of the most perilous effects of waterlogging is the spread of water and vector-borne diseases. Because of bad urban drainage, sewage from overflowing sewerage and latrines mix up with the rainfall runoff causing waterborne diseases. Moreover, stagnant water surrounding the houses provide a breeding ground for flies and mosquitoes leading to the spread of vector-borne diseases like malaria, dengue, and chikungunya. Bangladesh recently faced a severe dengue outbreak. DGHS (2019) data shows 44,986 patients in Dhaka were hospitalized, out of them 148 died. There is a dearth of research in Bangladesh on the correlations between water-logging and dengue fever. A study carried out by Salam (2018) in India's Delhi, however, found correlations between annual rainfall and dengue incidence. A comparable research to Salam's, if carried out in Dhaka, may potentially patterned the correlation.

Stagnant rainwater mixing with solid waste, litter, garbage, and other forms of pollutants damages trees, vegetation and aquatic resources. Common visible effects of urban pollution on surface water is the severe damage to the fish, and other aquatic animals and plants. When the logged water finally ends in the rivers or in the retention bodies, eroded sediments get deposited in the bottom, adversely affecting aquatic environment.

**4.4 Economic Problems:**

Owing to repeated water-logging, life span of roads and metalloid utility pipes decreases. Due to recurring construction and maintenance work, authorities have to bear huge cost. Creation of impervious surfaces by reason of construction work reduces lowering of the ground water, causing shortage of water, soil subsiding, and consolidation snags (Mowla, 2013). Also, a common case in Dhaka is that because of leakage or corrosion in supply water pipes, polluted rain water intrudes in them. Contaminated supply water critically exposes the users to various health hazards such as cholera, bacterial and protozoal diarrhea, influenza, hepatitis A and E, and typhoid fever. According to icddr,b (2014, pp. 11-17), such infections incur an "economic cost of at least US$169m a year in Bangladesh… nearly 6 million people are pushed into poverty each year in Bangladesh because of healthcare costs."

Rainwater often enters houses, damages household goods and stored food grains. Prolonged rainwater stay in the house dampens floor and wall, leading to direct financial cost. Waterlogging also damages income potentials as activities in markets, shops, and other financial enterprises receives deadly hits. Traffic jam as a result of waterlogging renders time, productivity and economic losses.

## 7. Remedial Policy Options

Rapid population expansion and subsequent unplanned urbanization are leading to illicit encroachment in the waterbodies and natural drainage system. Real estate and private developer activities, such as unlawful grabbing and filling of water retention bodies for housing purpose, ill-managed construction material dumping, are grossly violating the Wetland Conservation Act and derailing Dhaka's Detailed Area Plan (DAP). Such activities are resulting in the annihilation of natural drainage and water retention facilities, causing unprecedented water-logging. Only an appropriate set of remedial policy options can address this grave issue. This study prescribes the following steps:

### a. Natural Drainage System Recovery:

To ensure the ecological balance, a megacity must maintain at least 25 percent wetland; unfortunately, Dhaka has less than 10 percent (Khan cited in Tawhid, 2012). Concerned authorities such as BWDB, DNCC, DSCC, DWASA, RAJUK, etc. should take appropriate measures, and some of them are as follows:

- RAJUK needs to stop permitting the construction of buildings in low and wetlands.
- DWASA, BWDB, DCC (North and South) should work closely with other relevant authorities to identify a clear definition of water bodies, threatened to be filled out.
- All real estate and housing developers must be forced to follow the DMDP rules and regulations strictly. The DMDP demarcated flood retention areas should be left untouched for its respective uses.
- Recently, authorities have taken praiseworthy initiatives to recover khals and lakes from the encroachers (The Daily Star, 2017). However, experiences show that the encroachers always wait for the opportunity to resume their violation following the recovery. Strict imposition of post-recovery monitoring can stop them from resuming. Particularly, vigorous enforcement of laws and usage of the Wetland Conservation Act as an instrument can fend off the violators. If necessary, the Act should be amended.

### b. Waste Management:

Major Hurdles: Legal, Political, Financial, and Institutional

Lately, in Bangladesh, the burden of waste collection and disposal has befallen on the shoulder of the city corporations and the municipalities almost one-sidedly. With limited technical and financial capacities, Dhaka's City Corporations (North & South) find the waste management extremely painful. Unprecedented population density, explosive urban growth, careless littering practice made the cleaning of the city streets and drains nearly impossible. Bangladesh do not have any independent waste management law. The only legal instruments are the Bangladesh Environmental Conservation Act 1995 and Municipal Ordinance 1983 (amended in 1999). As ordinances do not strictly forbid littering, people dispose their wastes hither and thither and do not dispose them in the designated places or bins. The city urgently needs a new comprehensive and stringent solid waste management legislation, covering, a. the indexing of wastes in terms of the hazard scale, b. enforcing proper solid waste management procedures, c. introducing strict punishment against littering, so that people check themselves from throwing solid waste hither and thither. Enacting any laws in the National Parliament is tricky and lengthy. Nevertheless, adequate logical arguments to back it up can speedily overcome such political hurdles.

DNCC and DSCC combinedly hire approximately 3,000 temporary cleaners daily (Tawhid, 2012). These hiring is based on "no work – no payment," rendering job-insecurity and inefficient solid waste management. To improve the situation, they should revise the hiring method, job-duration and the remuneration. They must receive proper training, medical attention, as well as gloves and masks as protections.

As of today, modern scientific and technological methods are not applied in Dhaka's waste management. Therefore, most of the wastes are being dumped in the nearby water bodies and low lands.

Community Based Waste Management (CBWM)

DNCC and DSCC with their limited capacity cannot cope with Dhaka city's everyday solid waste. In many areas, community-based waste management has been introduced as an alternative. For instance, garbage pickup and disposal in Mirpur, Shaymoli, Kolabagan, Kathalbagan and Uttara has seen some success. Several non-governmental organizations are supporting this type of management. CBWM should receive strong sponsorship from the City Corporations. However, several matters need to be taken under consideration for an effective CBWM. Firstly, authorities should designate a convenient dumping location for the garbage collectors and should ensure efficient removal service for the garbage. Secondly, environment friendly waste disposal should replace the mass waste dumping in wetlands and water bodies; controlled incineration can be an option. Thirdly, the concerned authorities can introduce efficient recycling of the collected household waste.

**c. Improvement of drainage system:**

DWASA has prepared separate master plans on Dhaka's sewerage and drainage system in 2012 and 2016 respectively. Dhaka residents are yet to see any impact of those studies. However, the reports put forward several interesting proposals:

1. Incorporation of Basin Development Factor,
2. Comprehensive Drainage Development Plan,
3. Establish "Right-of Way,"
4. Green City Concept.

While the establish right-of way entails recovery from encroachments by establishing legal jurisdictions and the green city concept is still in the development face, the following subsections critically assess the first two proposals.

Incorporation of "Basin Development Factor": how realistic?

Considering Dhaka's S shape urbanization development, it is necessary to extend the watershed capacity. The drainage development masterplan fleetingly mentioned the incorporation of "Basin Development Factor (BDF)" to address the issue, without any proper execution strategy. In Texas, a study used more than 100 watersheds to generate estimations of peak discharge characteristics and unit hydrograph timing characteristics (Sundar et al. 2009). The objective of the study was to check whether BDF can accurately estimate the urban hydrologic response or not. Findings were discouraging-- BDF cannot make estimates worse but does not improve estimates either. This is not that the paper rejects BDF, but further in-depth feasibility study of BDF is required before implementing such an ambitious and costly project whimsically.

Comprehensive drainage development plan: impressive but execution required

The master plan provides a candid discussion on drainage development in Dhaka. The ideas and designs laid out in the report are impressive. It has been treated as a basic document for storm-water/manmade drainage network plan for Dhaka Area Plan, 2016-2035 (Dhaka Tribune, 2019 February). The report prescribes a number of recommendations including DWASA in collaboration with city corporations, placing proper drain outlets and cleaning them regularly, installing pipes under main roads and lanes in accordance with the water volume, placing sufficient drainage lines to cope with the heavy rainfall, building drains maintaining geographical locations, finishing digging without delay, recovering roads immediately after digging, removing garbage instantly after picking them from drains, reclaiming encroached waterbodies, installing separate sewerage line in every zone, not piling up construction materials on the road, cleaning up box culverts, maintaining open marshlands and lowlands, and constructing new water reservoirs. Other recommendations are: setting up new or expanding drainage infrastructure, maintaining saucer drains, catch pits, pipe drains, open drains, box culverts, storm sewer lines and, installing pumping stations and big water reservoirs.

Dhaka residents had seen many plans before but saw very little tangible improvement of the water-logging problem. The master plan is now nearly four years old. If the concerned authorities are truly sincere about addressing the waterlogging issue, they should execute the plans rather than carry forward the ideas from one document to another.

### d. Awareness raising against drain closing:

Strong awareness campaign should be introduced to inform the Dhaka residents about the aftereffects of natural drainage filling. People living nearby, occupy and pollute it day by day. Some people are illiterate, some are greedy, and others are careless. So, the awareness campaigns should be coupled with strong punishments for littering. Engaging NGOs can help the cause.

### e. Vegetation and aquaculture:

Water lily, lotus, ipil-ipil and other aquatic plants can be cultivated in Dhaka's waterbodies. Because of their deep roots and huge nutrient absorption capacity, they can clear the polluted waters and restrict eutrophication (Mowla, 2008). Fish cultivation is also a good way to keep the water clean. Eco friendly fishes such as Sarputi, Rajputi, Rui, Silver Curp, Grass Curp eat upper level foods and wastage in the water such as rotten leaves, organic materials, insects and thus purifies water (Mowla, 2013). Another technique of resuscitating lost water-habitat is thick vegetation along major water channels. Planting hardy species such as Ashoke, Mahua, Hijal, Keora, Krishnachura, Kadam, Jarul, Shiuli etc. can reduce hydrological pollution, soil erosion, sedimentation and eutrophication (Mowla, 2008).

### f. Replication of Best Practices and Possible Collaborative Approaches:

Bangladesh can learn from the best practices from other nations. For instance, Netherlands has a proud history of flood management. Praiseworthy Dutch projects such as 'room for rivers' can be a big inspiration. Replication of such project through collaborative efforts between the Dutch government and the Bangladesh government opens a more meaningful window to address Dhaka water-logging.

## 8. Conclusion:

Unplanned development is the key reason behind Dhaka's waterlogging. Owing to the unplanned development, the storm water drainage systems, are facing illegal encroachment, filling up, causing diversion and obstruction to the natural flow of water. Because of this obstruction, the city is exposed to irksome waterlogging, particularly during monsoon. Waterlogging incurs substantial direct and indirect adverse social, physical, economic and environmental costs. Such costs during the downpours indicate the seriousness and necessity for corrective government responses. Lack of coordination and transparency are prevalent among the concerned authorities. Negligence and carelessness of the city corporation staff and cleaners, careless littering by the citizens and procrastination in drainage construction and maintenance work are some of the key reasons behind the city's waterlogging problems.

Sustainable drainage could be key to a sustainable development as it decreases the impacts would otherwise transpire due to heavy downpour and water runoff. Because the works are interrelated, if the drainage system works properly, the utility services could also work properly. An effective synchronization and collaboration between concerned authorities is important for an operative drainage system to improve the city's waterlogging problem.

Annexure:

*Table 6: Focused Group Discussions and Personal Interviews Information.*

| Technique | Respondents | Affiliation/ Area |
|---|---|---|
| Informal Interview | 1. Executive Engineer (Mollah Md. Nuruzzaman) | Zone 4, Dhaka North City Corporation (DNCC) |
| | 2. Professor (Dr. Shah Jahan Mandal) | Institute of Water and Flood Management, BUET |
| | 3. Professor (Dr. Md. Shakil Akter) | Urban and Regional Planning, BUET |
| | 4. Ward Counsellor (Humayun Rashid Jonny) | Ward -14, Dhaka |
| Focused Group Discussions | 5. Residents (8) | Kazipara, Mirpur, Dhaka (DNCC) |
| | 6. Residents (8) | Moulivirtek, Khilgaon, Taltola, Dhaka (DSCC) |
| | 7. Residents (8) | Basabo, Sabujbag, Dhaka (DSCC) |

*Figure 1: Dhaka's Natural Drainage System and Elevation (Source: DWASA).*

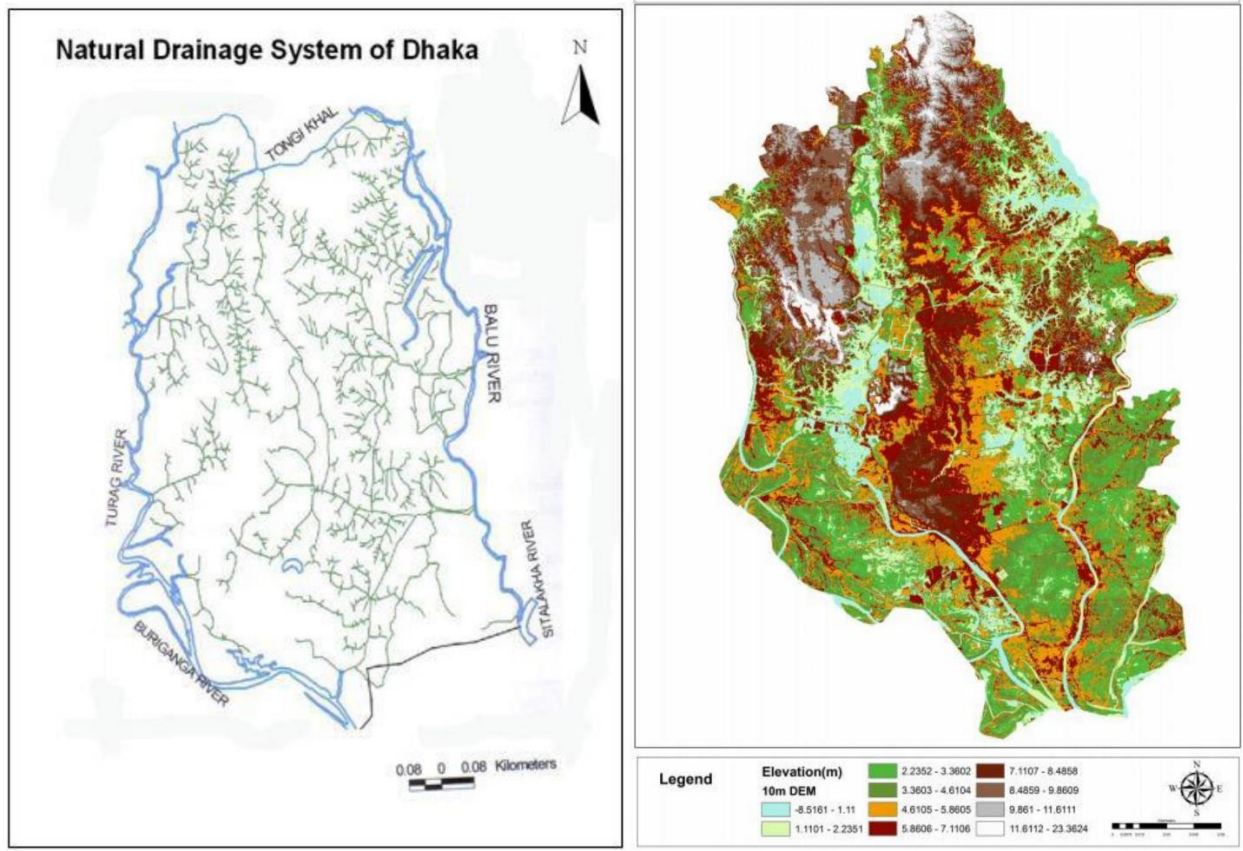

*Figure 2: Retention Pond (Source: DNCC, Zone 4, 2019).*

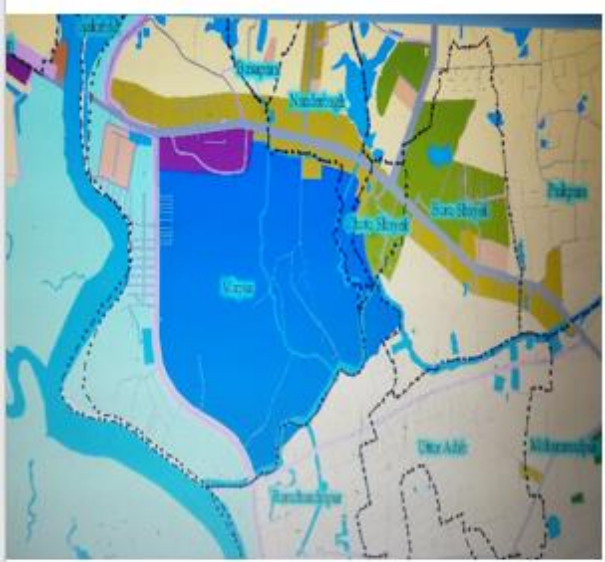
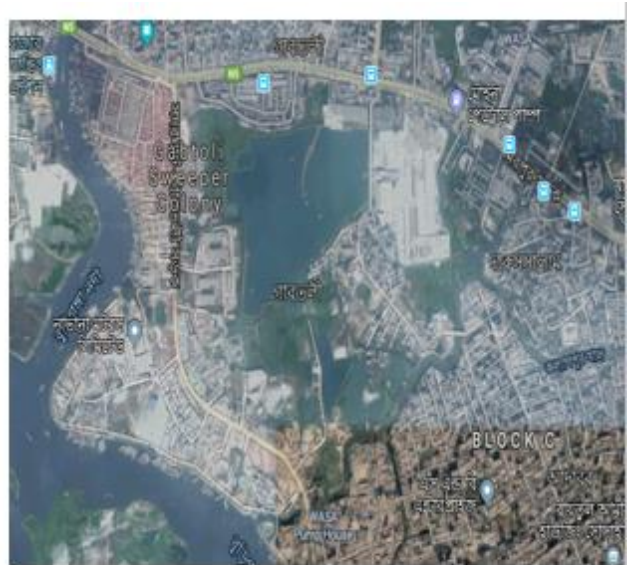

Retention Pond at Mouja Map 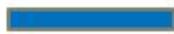   Retention Pond at Google Satellite Pic

*Figure 3: Dhaka's Storm Water Drainage System (DWASA, 2012).*

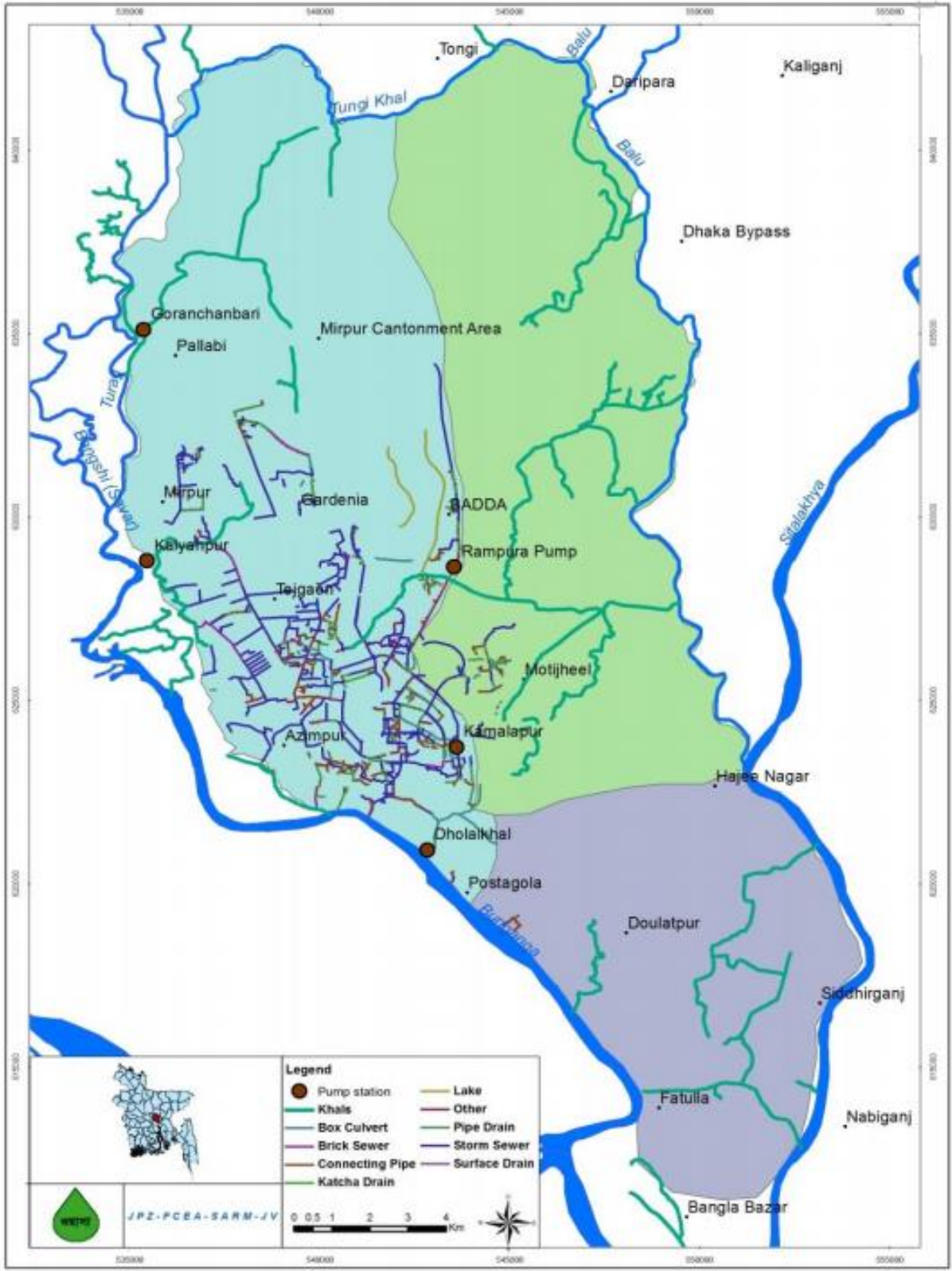

*Figure 4: DNCC Zone 4- Base Map- Roads (Source: DNCC, Zone 4 Office, 2019).*

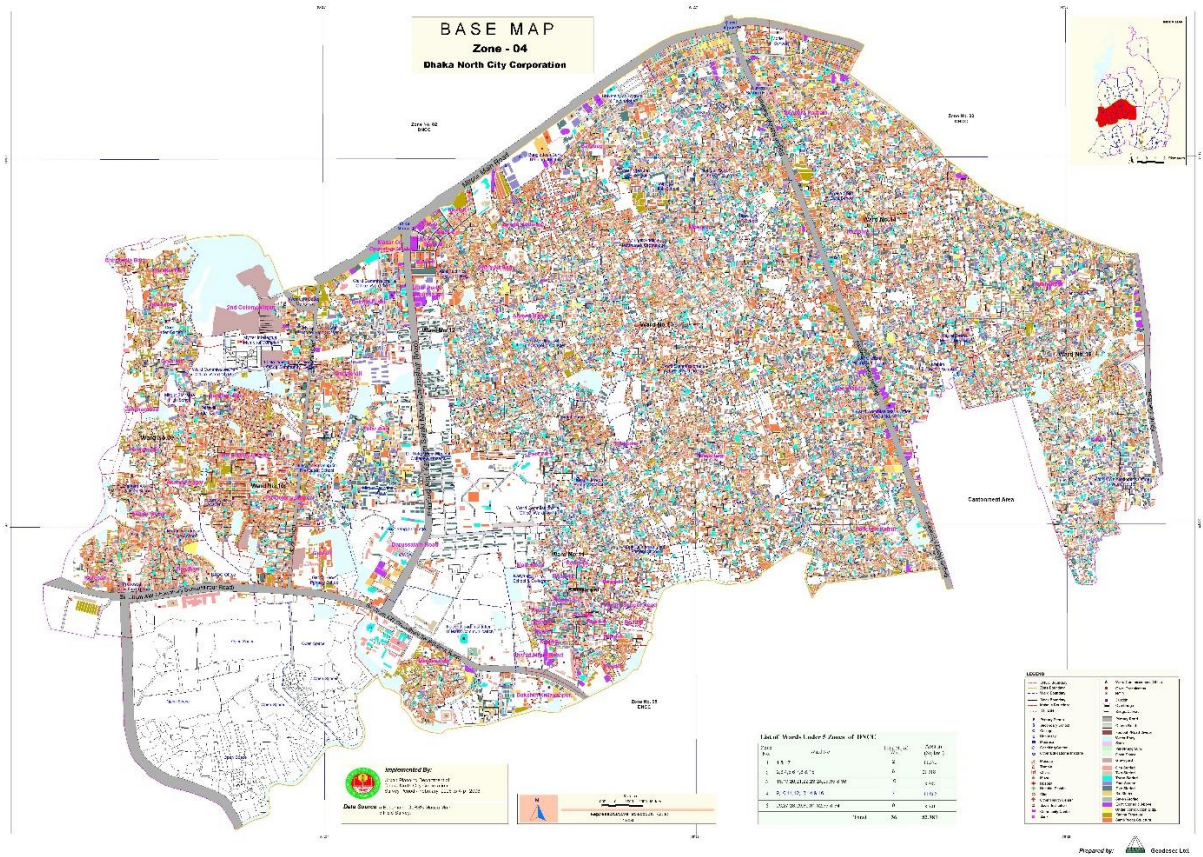

*Figure 5: Wastes in Khals (Picture 1) and Encroached Khals (Picture 2) (Source DNCC, Zone 4 Office, 2019)*

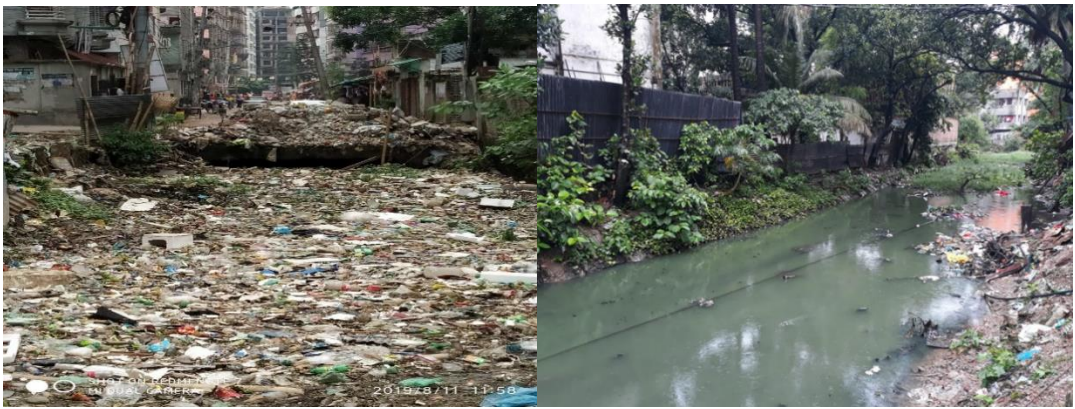

*Figure 6: Proposed drainage facilities of Dhaka region (RAJUK, 2015).*

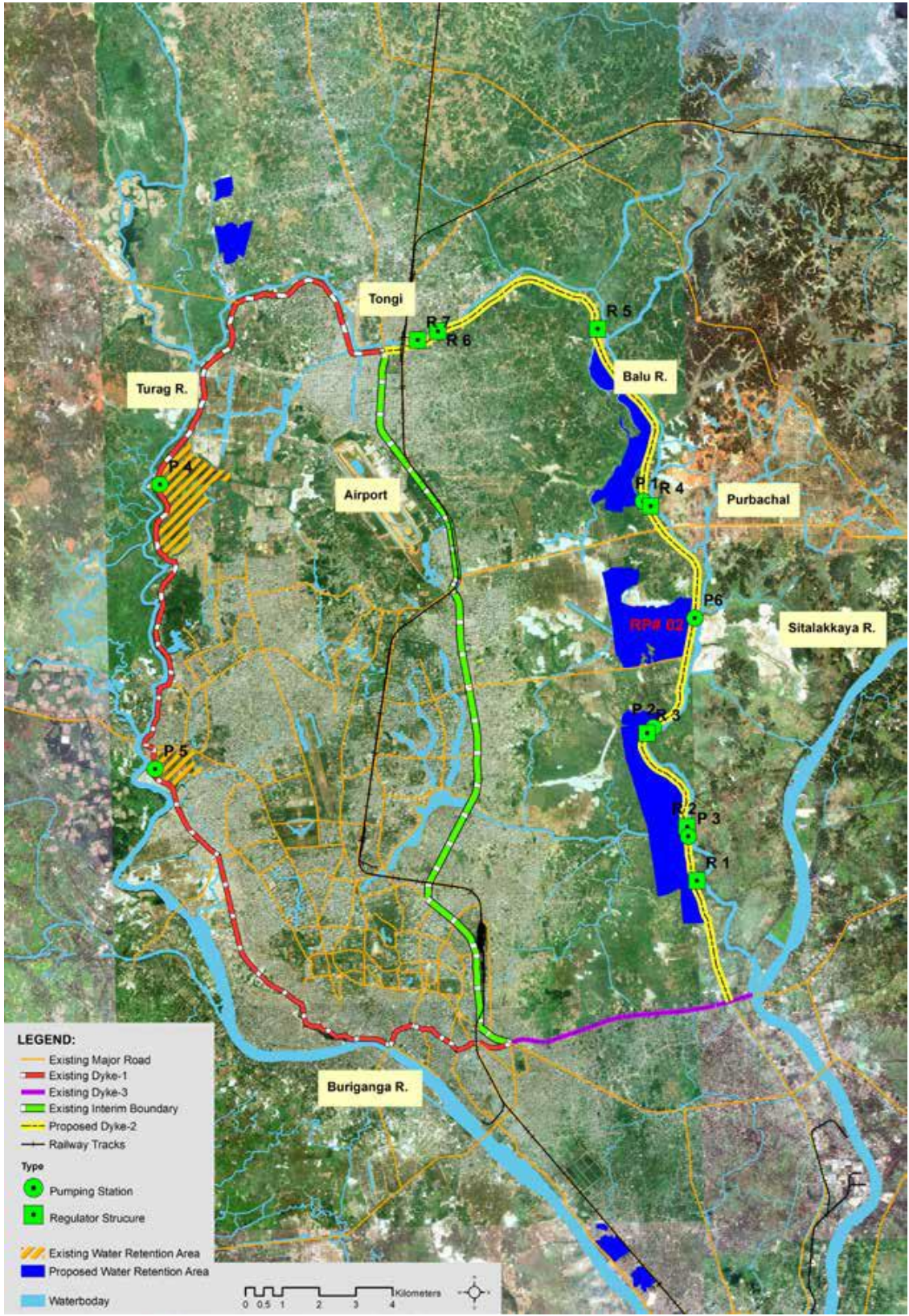